\documentclass{article}
\usepackage{ijcai26}

\usepackage{times}
\usepackage{soul}
\usepackage{pifont}
\usepackage{url}
\usepackage{array}
\usepackage{multirow}
\usepackage{amsmath}
\usepackage{makecell}
\usepackage[hidelinks]{hyperref}
\usepackage{adjustbox}
\usepackage[utf8]{inputenc}
\usepackage[small]{caption}
\usepackage{graphicx}
\usepackage{amsthm}
\usepackage{booktabs}
\usepackage{algorithm}
\usepackage{algorithmic}
\usepackage[switch]{lineno}
\usepackage{tcolorbox}
\usepackage{adjustbox}

\usepackage{tikz}
\usetikzlibrary{
    arrows.meta,
    shapes.geometric, % For rounded corners and other shapes
    positioning
}
\usepackage{capt-of}
\usepackage[edges]{forest}

\tikzset{
  my-box/.style={rectangle, draw=hidden-black, rounded corners, text opacity=1, minimum height=1.5em,
    minimum width=5em, inner sep=2pt, align=center, fill opacity=.5,},
  leaf/.style={my-box, minimum height=1.5em, fill=yellow!32, text=black, align=left, font=\normalsize,
    inner xsep=5pt, inner ysep=4pt, text width=45em,},
  leaf2/.style={my-box, minimum height=1.5em, fill=purple!27, text=black, align=left, font=\normalsize,
    inner xsep=5pt, inner ysep=4pt,},
  leaf3/.style={my-box, minimum height=1.5em, fill=hidden-blue!57, text=black, align=left, font=\normalsize,
    inner xsep=5pt, inner ysep=4pt,},
  leaf4/.style={my-box, minimum height=1.5em, fill=green!14, text=black, align=left, font=\normalsize,
    inner xsep=5pt, inner ysep=4pt,},
  leaf5/.style={my-box, minimum height=1.5em, fill=orange!16, text=black, align=left, font=\normalsize,
        inner xsep=5pt, inner ysep=4pt,},
}

\tcbset{
  takeawaybox/.style={width=\linewidth, top=8pt, bottom=4pt, colback=hidden-yellow, colframe=black,
    colbacktitle=black, fonttitle=\bfseries, enhanced, center title, attach boxed title to top left={yshift=-0.1in,xshift=0.15in},
    boxed title style={boxrule=0pt,colframe=white,},
  }
}
\newtcolorbox{TakeawayBox}[2][]{takeawaybox,title=#2,#1}

\usetikzlibrary{arrows.meta,shapes,positioning,shadows,trees}
\definecolor{hidden-red}{RGB}{205, 44, 36}
\definecolor{hidden-blue}{RGB}{194,232,247}
\definecolor{hidden-orange}{RGB}{243,202,120}
\definecolor{hidden-green}{RGB}{34,139,34}
\definecolor{hidden-pink}{RGB}{255,245,247}
\definecolor{hidden-black}{RGB}{20,68,106}
\definecolor{purple}{RGB}{144,153,196}
\definecolor{yellow}{RGB}{255,228,123}
\definecolor{hidden-yellow}{RGB}{255,248,203}
\definecolor{tkcolor}{RGB}{224,223,255}
\definecolor{hidden-draw}{RGB}{128,128,128}
\definecolor{darkblue}{rgb}{0, 0.40, 0.75}
\definecolor{colorEssential}{HTML}{0D47A1}
\definecolor{colorCommon}{HTML}{1E88E5}
\definecolor{colorNiche}{HTML}{BDBDBD}

\title{Multimodal Emotion Recognition with Large Language Models}
\author{
    Hongrui Zhang$^{1,2}$\thanks{Equal contribution.}\and
    Daiqing Wu$^{1}$\footnotemark[\value{footnote}]\and
    Yangyang Li$^{3}$\and
    Kuien Liu$^{3}$\and
    Yuhui Wang$^{3}$\and \\
    Yu Zhou$^{4}$\and
    Sicheng Zhao$^{1}$\thanks{Corresponding author.}
    \affiliations
    $^1$Department of Psychological and Cognitive Sciences, Tsinghua University, Beijing, China\\
    $^2$School of Computer Science and Technology, Harbin Institute of Technology, Weihai, China\quad
    $^3$Academy of Cyber, Beijing, China\\
    $^4$College of Computer Science, Nankai University, Tianjin, China\\
    \emails
    smilingweeping@gmail.com,\ 
    schzhao@tsinghua.edu.cn
}

\begin{document}
\maketitle
% Abstract
% \begin{abstract}
% In many practical applications, directly applying general multimodal large language models (MLLMs) to multimodal emotion recognition (MER) tasks is challenging due to their limitations in both capturing subtle perceptual cues and in the nuanced interpretation of emotions, which can lead to inaccurate emotional recognition or even hallucinations. To address these shortcomings, significant research efforts have focused on enhancing the ability of MLLMs in the MER domain. In this paper, we systematically review the key challenges in MER and provide a comprehensive comparison of contemporary approaches from multiple perspectives. Finally, we discuss promising future research directions aimed at advancing MLLMs for more accurate and robust emotion recognition.
% \end{abstract}
\begin{abstract}
%Multimodal Emotion Recognition (MER) focuses on identifying and interpreting human emotions from various modalities, including text, speech, and visual cues. It has a wide range of applications, from human-computer interaction and healthcare to marketing and social media analytics. Recent advancements in MER are characterized by a paradigm shift from small-scale, task-specific models to multimodal large language models (MLLMs). The shift to MLLMs represents a solid step towards generalizable emotional intelligence, pushing the boundaries of MER. However, with these new opportunities come new challenges, including affective data scarcity, multimodal affective gap, and affective interpretation opacity. In response, we review three corresponding approaches: affective data augmentation, multimodal affective representation, and multimodal affective reasoning. In this paper, we provide an overview of these advancements, discussing the key developments, emerging trends, and open challenges in the evolving field of MER.
Multimodal Emotion Recognition (MER) focuses on identifying and interpreting emotions from modality-compound inputs. Closely mirroring human cognitive processes in real-world environments, MER has drawn substantial attention from both academia and industry. Recently, a paradigm shift has been unveiled in MER, from leveraging small-scale, task-specific models to Large Language Models (LLMs). We refer to the latter as the MER-with-LLMs paradigm, which offers unprecedented generality, spurring numerous empirical attempts, even alongside speculation about LLMs' potential to achieve general emotional intelligence. However, with these new opportunities come new challenges, including the scarcity of emotionally annotated data, the affective gap both within and across modalities, and the opacity of affective interpretation. To systematically review existing research and guide future exploration, this paper categorizes prior works according to their focus on addressing these challenges into three directions: Affective Data Augmentation, Multimodal Affective Representation, and Multimodal Affective Reasoning. By thoroughly tracing the development, emerging trends, and remaining issues within each direction, this paper aims to provide a clear academic map of the MER-with-LLMs paradigm and foster its structured advancement.
\end{abstract}
% Emotion recognition (MER) has witnessed a paradigm shift, transitioning from small-scale models to large multimodal language models (MLLMs). This transition has significantly enhanced the flexibility and scalability of MER, marking a solid step toward achieving generalizable emotional intelligence. While these advancements present new opportunities, they also introduce unique challenges. This survey systematically organizes three critical challenges in the field: affective data scarcity, the multimodal affective gap, and affective interpretation opacity. In response, we review three corresponding approaches: affective data augmentation, multimodal affective representation, and multimodal affective reasoning. By addressing these challenges, the field moves closer to developing robust, scalable systems for emotional intelligence across diverse contexts.
% Introduction
\section{Background and Challenges} \label{sec:introduction}
Emotion is an integral component of human daily experiences, significantly influencing individuals’ communication, decision-making, and behavior. In real-world settings, emotions are expressed and perceived through multiple modalities, including language, speech, facial expressions, and gestures~\cite{label-eff-survey,MIR-gesture}. As a result, Multimodal Emotion Recognition (MER) has emerged as a critical research area, focusing on enabling models to comprehend emotions from modality-compound inputs as humans do. %typically through classification.
To achieve this, traditional research primarily relies on small-scale, task-specific models~\cite{small-scale1,small-scale2}. Despite achieving satisfactory performance, they suffer from strong dependencies on predefined input domains and output spaces, which make them less effective in practical applications that require dynamism and flexibility.
\begin{figure}[t]
    \centering
    \includegraphics[width=1\linewidth]{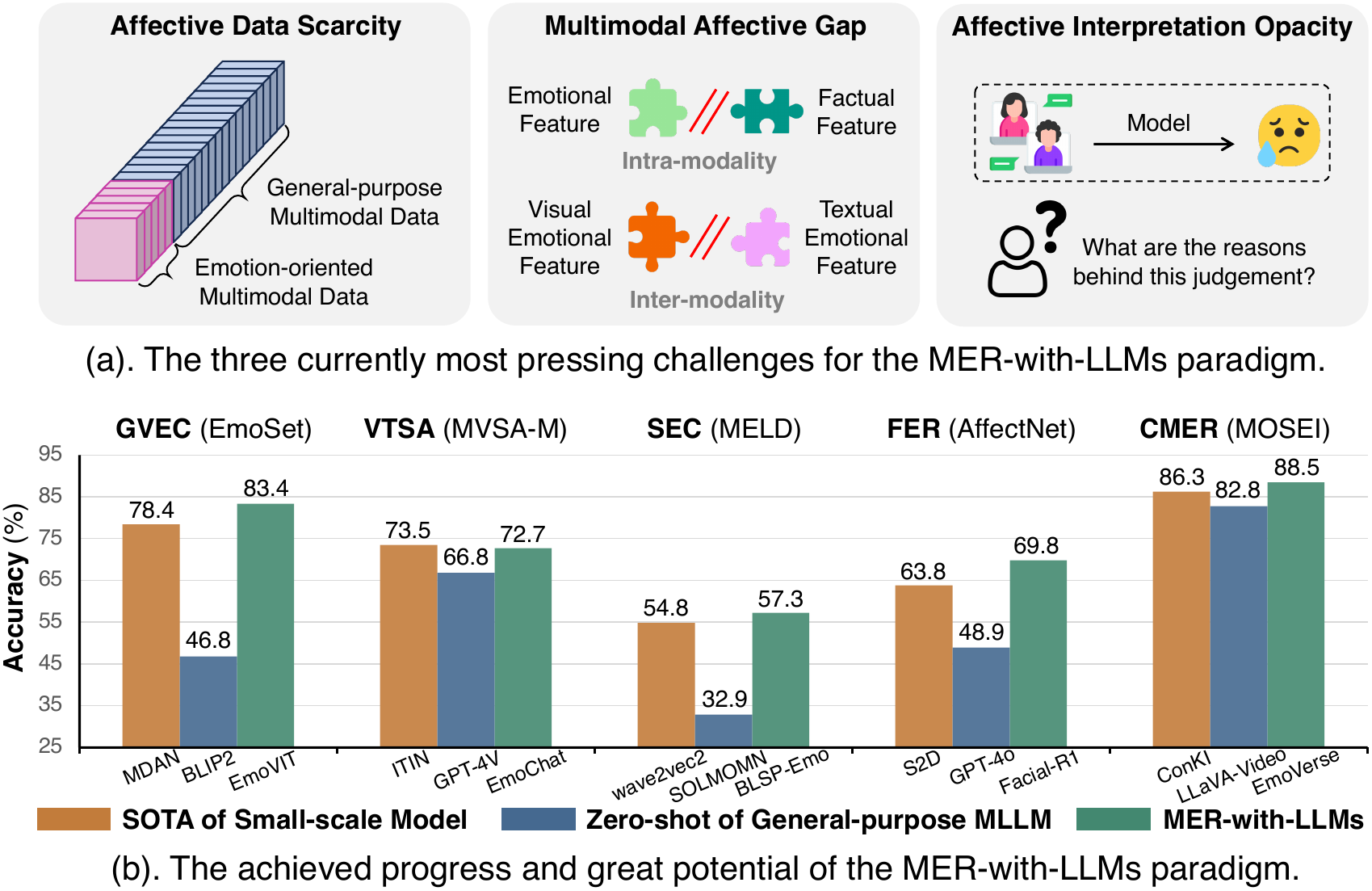}
    \caption{The MER-with-LLMs paradigm faces three major challenges, along with significant potential and opportunities.}
    \label{fig:motivation}
\end{figure}
\tikzstyle{my-box}=[
    rectangle,
    draw=hidden-draw,
    rounded corners,
    text opacity=1,
    minimum height=1.5em,
    minimum width=5em,
    inner sep=2pt,
    align=center,
    fill opacity=.5,
]

\tikzstyle{level1-style}=[
    my-box,
    fill=gray!5,
    text=black,
    font=\scriptsize,
    inner xsep=2pt,
    inner ysep=3pt,
]

\tikzstyle{level2-style}=[
    my-box,
    fill=gray!10,
    text=black,
    font=\scriptsize,
    inner xsep=2pt,
    inner ysep=3pt,
]

\tikzstyle{level3-style}=[
    my-box,
    fill=gray!15,
    text=black,
    font=\scriptsize,
    inner xsep=2pt,
    inner ysep=3pt,
]

\tikzstyle{leaf}=[
    my-box, 
    minimum height=1.5em,
    fill=yellow!32, 
    text=black, 
    align=left,
    font=\scriptsize,
    inner xsep=2pt,
    inner ysep=4pt,
]
\tikzstyle{leaf2}=[
    my-box, 
    minimum height=1.5em,
    fill=purple!27, 
    text=black, 
    align=left,
    font=\scriptsize,
    inner xsep=2pt,
    inner ysep=4pt,
]
\tikzstyle{leaf3}=[
    my-box, 
    minimum height=1.5em,
    fill=hidden-blue!57, 
    text=black, 
    align=left,
    font=\scriptsize,
    inner xsep=2pt,
    inner ysep=4pt,
]
\tikzstyle{leaf4}=[
    my-box, 
    minimum height=1.5em,
    fill=green!14, 
    text=black, 
    align=left,
    font=\scriptsize,
    inner xsep=2pt,
    inner ysep=4pt,
]
\tikzstyle{leaf5}=[
    my-box, 
    minimum height=1.5em,
    fill=orange!16, 
    text=black, 
    align=left,
    font=\scriptsize,
    inner xsep=2pt,
    inner ysep=4pt,
]

\begin{figure*}[t]
    \centering
    \resizebox{\textwidth}{!}{
        \begin{forest}
            forked edges,
            for tree={
                grow=east,
                reversed=true,
                anchor=base west,
                parent anchor=east,
                child anchor=west,
                base=left,
                font=\scriptsize,
                rectangle,
                draw=hidden-draw,
                rounded corners,
                align=left,
                minimum width=4em,
                edge+={darkgray, line width=1pt},
                s sep=3pt,
                inner xsep=2pt,
                inner ysep=3pt,
                ver/.style={rotate=90, child anchor=north, parent anchor=south, anchor=center},
            },
            where level=1{text width=13em,font=\scriptsize,level1-style}{},
            where level=2{text width=12.4em,font=\scriptsize,level2-style}{},
            where level=3{text width=7em,font=\scriptsize,level3-style}{},
            where level=4{text width=6.1em,font=\scriptsize,}{},
            [
                MER-with-LLMs, ver
                [
                     Affective Data Augmentation~(\S\ref{sec:augmentation})
                    [
                        % Training-free Sample Configuration, font=\fontsize{6.7}{8}\selectfont,
                        Training-free Sample Configuration, font=\scriptsize,
                        [
                            \textit{e.g.,}
                             NarraCap~\cite{EMOTIC}{,}
                            SoV~\cite{SoV}{,}
                            SoVTP~\cite{SoVTP}{,}
                            In-Context Learning~\cite{ICL}
                            , leaf3, text width=40.7em
                        ]
                    ]
                    [
                        Emotion Data \\ Annotation, align=center, text width=5.6em
                        [
                            Dataset Engineering
                            [
                                \textit{e.g.,}
                                EmoVIT~\cite{EmoVIT}{,}
                                VEC-CoT~\cite{EmoCaliber}{,}
                                FABA-Instruct~\cite{EmoLA}{,}
                                DEEMO~\cite{DEEMO}{,} \\
                                EmoCause~\cite{EmoDETective}{,}  
                                MEC$^{4}$~\cite{M3F}{,}
                                MESC~\cite{MESC}{,} 
                                MERR~\cite{Emotion-LLaMA}{,} \\
                                MER-Caption~\cite{AffectGPT}{,}
                                AMT~\cite{EmoVerse}
                                , leaf3, text width=39em
                            ]
                        ]
                        [
                            Benchmark Construction
                            [
                                \textit{e.g.,}
                                VECBench~\cite{EmoCaliber}{,}
                                EEmo-Bench~\cite{EEmoBench}{,}
                                MVEI~\cite{MVEI}{,}
                                MM-BigBench~\cite{MMBigBenchEM}{,} \\
                                FaceBench~\cite{FaceBench}{,}
                                EIBench~\cite{EIBench}{,}
                                MTMEUR~\cite{MTMEUR}{,}
                                Hi-EF~\cite{Hi-EF}{,} \\
                                CA-MER~\cite{MoSEAR}{,}
                                MER-UniBench~\cite{AffectGPT}{,}
                                OV-MERD~\cite{OVMER}{,}
                                EmotionHallucer~\cite{EmotionHallucer}
                                , leaf3, text width=39em
                            ]
                        ]
                    ]   
                ]
                [
                    Multimodal Affective Representation~(\S\ref{sec:representation})
                    [
                        Perceptual Emotion Mapping
                        [
                            \textit{e.g.,}
                            SEPM~\cite{SEPM}{,}
                            MERMAID~\cite{MERMAID}{,}
                            StimuVAR~\cite{StimuVAR}{,}
                            EmoDETective~\cite{EmoDETective}{,} \\
                            ExpLLM~\cite{ExpLLM}{,}
                            EmoLA~\cite{EmoLA}{,}
                            FaceInsight~\cite{FaceInsight}{,}
                            BLSP-Emo~\cite{BLSP-Emo}{,} \\
                            SECap~\cite{SECap}{,}
                            EmoVerse~\cite{EmoVerse}{,}
                            Emotion-LLaMA~\cite{Emotion-LLaMA}{,}
                            EmoChat~\cite{EmoChat}
                            , leaf5, text width=40.8em
                        ]
                    ]
                    [
                        Multimodal Emotion Coordination
                        [
                            \textit{e.g.,}
                            EmoVIT~\cite{EmoVIT}{,}
                            Omni-SILA~\cite{Omni-SILA}{,}
                            FEALLM~\cite{FEALLM}{,}
                            Omni-Emotion~\cite{Omni-Emotion}{,} \\
                            AffectGPT~\cite{AffectGPT}{,} 
                            M$^{3}$F~\cite{M3F}{,}
                            EMO-LLaMA~\cite{EMO-LLaMA}
                            , leaf5, text width=40.8em
                        ]
                    ]
                ]
                [
                   Multimodal Affective Reasoning~(\S\ref{sec:reasoning})
                    [
                        Emotion Explanation and Hallucination, align=left
                        [
                            \textit{e.g.,}
                            % EMER~\cite{EMER}{,}
                            MulCoT-RD~\cite{MulCoT-RD}{,}
                            Facial-R1~\cite{Facial-R1}{,}
                            AlignCap~\cite{AlignCap}{,}
                            Affective-CoT~\cite{Affective-CoT}{,} \\
                            PEP-MEK~\cite{EmotionHallucer}{,}
                            ERV~\cite{ERV}{,}
                            FIRES~\cite{FIRES}{,} 
                            % MoSEAR~\cite{MoSEAR}{,} 
                            MultiMood~\cite{MultiMood} 
                            , leaf4, text width=40.8em
                        ]
                    ]
                    [
                        Subjective Emotion Reasoning, align=left
                        [
                            \textit{e.g.,}
                            % OV-MER~\cite{OVMER}{,}
                            EmoCaliber~\cite{EmoCaliber}{,}
                            Agent-MER~\cite{Agent-MER}{,}
                            AffectGPT-R1~\cite{AffectGPT-R1}
                            , leaf4, text width=40.8em
                        ]
                    ]
                ]
            ]
        \end{forest}
    }
    % \caption{Taxonomy of the MER-with-LLMs paradigm. We organize existing works into three branches according to their main focused challenges: Affective Data Augmentation for Affective Data Scarcity, Multimodal Affective Representation for Multimodal Affective Gap, and Multimodal Affective Reasoning for Affective Interpretation Opacity. Representative works are listed on the leaves.}
    \caption{Taxonomy of the MER-with-LLMs paradigm. We categorize existing works into three branches based on their primary challenges, with representative works listed on the leaves.}
    \label{fig:taxonomy}
\end{figure*}
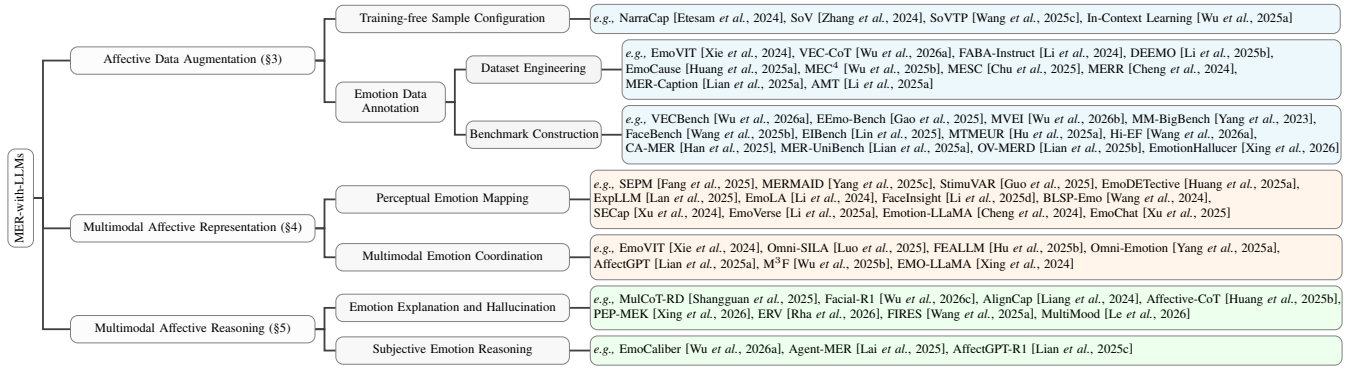

The emerging Large Language Models (LLMs) present a promising avenue for addressing these limitations \cite{MER-Survey}. Large-scale generative pre-training equips LLMs with strong instruction-following capabilities, which are inherited by Multimodal LLMs (MLLMs), giving rise to a new \textbf{MER-with-LLMs} paradigm. By formulating MER as an LLM-centric autoregressive process, this paradigm enables unified handling of inputs spanning multiple domains and modalities, as well as producing diverse instruction-conditioned outputs. It even deepens the MER task itself, extending classification toward explanation~\cite{EMER}. Collectively, these fundamental advantages have driven a paradigm shift, reflected by a rapidly growing body of recent work on MER-with-LLMs.

% Despite these advances, several challenges remain in leveraging MLLMs for MER and affective computing tasks. Although MLLMs show impressive performance in general multimodal tasks, their zero-shot capabilities are still limited when applied specifically to MER tasks \cite{GPT-4V}. Additionally, general MLLMs face significant challenges, such as emotion stereotyping \cite{SeeingRace} and hallucinations \cite{EmotionHallucer}, which arise from inadequate training data and inherent model limitations. These issues hinder the seamless integration of emotion recognition into multimodal systems, thus impeding their practical application in affective domains. \\
% \indent Given these limitations, an increasing body of research has begun to analyze the key elements affecting the emotion prediction performance of MLLMs \cite{Analyzing} and focus on enhancing the affective understanding of MLLMs, aiming to improve their ability to accurately perceive and interpret human emotions across modalities. However, a comprehensive survey that systematically compiles the progress in this field is still lacking. To this end, we propose a comprehensive taxonomy categorizing methodologies into three primary challenges:\\
Alongside this trend, new challenges have also arrived. A line of studies~\cite{GPT-4V,ICL} has reached a consensus that, under zero-shot inference, MLLMs often fail to achieve proficiency on MER comparable to that observed in other multimodal tasks. This deficiency is inherently rooted in the complexity of MER, which entails capturing high-level affective cues from multimodal inputs, modeling interactions among heterogeneous modalities, and integrating them to derive emotional conclusions. 
% To address these challenges, various optimization strategies with divergent emphases have been proposed and substantially advanced this paradigm, as quantitatively evidenced by Fig.~\ref{fig:motivation}(b). However, their optimization pathways remain intricate and diversified, and a systematic organization is still lacking. To promote a more structured and coherent development of this emerging paradigm, this paper provides a comprehensive review of the existing challenges, a taxonomy of current methods, and an in-depth discussion of recent progress and future directions. 
Following a progressive order, we categorize the currently most pressing challenges into three branches, as illustrated in Fig.~\ref{fig:motivation}(a): \\
\textbf{Affective Data Scarcity.} Data constitute the foundation of modern models. The underperformance of general MLLMs on MER underscores the importance of emotional data, making targeted optimization on such data a direct and intuitive remedy. However, the subjectivity of emotion perception often necessitates collaboration from multiple annotators for reliable annotation, substantially increasing labeling costs and constraining the scale of high-quality datasets. Moreover, datasets collected in different contexts are typically annotated under distinct and often incompatible emotion theories, further aggravating dataset fragmentation. These limitations compel existing approaches to either maximize the utilization of available data or construct new datasets. \\
%General MLLMs training datasets require large, diverse samples to enhance model generalization across tasks. In contrast, affective datasets require careful design to account for context, culture, and individual differences, with fine-grained annotations for emotional cues, particularly facial expressions and intonations, making them smaller and more complex.
\textbf{Multimodal Affective Gap.} The ``affective gap'' generally refers to the intra-modality misalignment between emotional and factual features~\cite{tpami_survey}, which challenges MLLMs in capturing meaningful affective cues. We extend this concept to cover an inter-modality setting, where it characterizes the heterogeneity and semantic discrepancy of emotional features across different modalities. Heterogeneity reflects inconsistencies in information density and expression patterns; for instance, text tends to be more compact and abstract, whereas images are more dispersed and concrete. Semantic discrepancy refers to the fact that each modality provides unique and sometimes conflicting cues. 
% Together, these factors substantially complicate the modeling of inter-modality synergy among emotional features. Overall, these gaps drive the exploration of dedicated affective cues capturing and interaction modeling.
Together, these factors complicate inter-modality emotional synergy modeling. Overall, these gaps motivate the exploration of dedicated affective cues extraction and interaction modeling.\\
%The complexity of emotional and factual characteristics arises from the heterogeneous nature of emotional cues across different modalities, such as facial expressions, voice tone, body language, and text. Each modality provides unique and sometimes conflicting emotional information, making it difficult to effectively capture and integrate these cues into a coherent emotional representation~\cite{SDRS}. Effectively capturing these heterogeneous cues and integrating them into a unified emotional representation requires models capable of discerning and reconciling these complex, sometimes contradictory, signals. The challenge lies not only in detecting emotional signals but also in understanding how they interact and contribute to the overall emotional state in multimodal contexts.
\textbf{Affective Interpretation Opacity.} Inherited from the habits of small-scale models, mainstream MER methods primarily focus on emotion recognition rather than explanation. 
% This emphasis results in the lack of decision transparency, which not only hinders iterative model improvement but also raises concerns about reliability in high-stakes applications, such as healthcare, emotional support, and education.
This emphasis results in a lack of decision transparency, which not only hinders iterative model improvement but also raises concerns about reliability in high-stakes and cross-domain applications, including healthcare, education, and finance.
Benefiting from the autoregressive generation process of LLMs, the MER-with-LLMs paradigm naturally supports step-by-step reasoning or explicit natural language explanations for emotional decisions, thereby giving rise to a new task: explainable MER. In this context, existing approaches aim to enhance model transparency and reliability, for example, by reducing hallucinations, eliciting confidence verbalization, and enabling open-vocabulary MER.
%Despite advancements in model performance, current MLLMs often struggle with transparency in the emotional prediction process. The lack of interpretability and clear decision-making paths in emotion recognition models undermines trust, especially in high-stakes applications such as healthcare, emotional support, and education. This opacity makes it difficult to assess how the model arrives at its predictions and raises concerns about its reliability and fairness. To ensure broader adoption and trust, it is crucial to develop methods that provide transparent and understandable explanations of how emotional predictions are derived, addressing issues such as hallucinations, emotional biases, and the challenges of Open-Vocabulary (OV) emotion recognition.

% To address these challenges, various optimization strategies with divergent emphases have been proposed and substantially advanced this paradigm, as quantitatively evidenced by Fig.~\ref{fig:motivation}(b).
To address these challenges, various optimization strategies with divergent emphases have been proposed, significantly advancing this paradigm, as illustrated in Fig.~\ref{fig:motivation}(b).
However, their optimization pathways remain intricate and diversified, and a systematic organization is still lacking. To promote a more structured development of this emerging paradigm, this paper provides a comprehensive review of the existing challenges, a taxonomy of current methods, and an in-depth discussion of recent progress and future directions. 
% Problem Formulation
\section{Problem Formulation and Taxonomy} \label{sec:challenge}
% Affective Dataset and Benchmark
\begin{table*}[tb]
\centering
\resizebox{\textwidth}{!}{
\begin{tabular}{@{}clcccccc@{}}
\toprule
& \textbf{Method} & \textbf{Modality} & \textbf{Sub-task} & \textbf{\#Sample} & \textbf{\#Labels} & \textbf{Annotators} & \textbf{Key Focus} \\
% DEEMO-NFBL is body language, columns is Emotions or Labels
\toprule
% \multicolumn{7}{c}{\textit{\textbf{Training Dataset}}} \\
% \midrule
\multirow{13}{*}{\rotatebox{90}{Training Dataset}}
& EmoVIT~\cite{EmoVIT} & I,T & GVEC & 51,200 & - & Model & Emotional Instruction \\
& VEC-CoT~\cite{EmoCaliber} & I,T & GVEC & 143,446 & - & Model & Structured Affective Reasoning \\
& FABA-Instruct~\cite{EmoLA} & I,T & FER & 19,877 & 7 & Model-led+Human-assisted & Face Analysis \\
& DEEMO-MER~\cite{DEEMO} & V,A,T & CMER & 2,060 & 2 & Model-led+Human-assisted & Identity Privacy \\
& DEEMO-NFBL~\cite{DEEMO} & V,A,T & CMER & 24,722 & 37 & Human & Identity Privacy \\
& EmoCause~\cite{EmoDETective} & V,A,T & CMER & 14,000 & 8 & Model-led+Human-assisted & Emotional Cause Understanding \\
& MEC$^{4}$~\cite{M3F} & V,A,T & CMER & 2,124 & - & Human & Emotional Cause Understanding \\
& MESC~\cite{MESC} & V,A,T & CMER & 28,762 & 7 & Human-led+Model-assisted & Emotional Support \\
& MERR-Coarse~\cite{Emotion-LLaMA} & V,A,T & CMER & 28,618 & - & Model & Multimodal Emotion Instruction \\
& MERR-Fine~\cite{Emotion-LLaMA} & V,A,T & CMER & 4,487 & - & Human-led+Model-assisted & Multimodal Emotion Instruction \\
& MER-Caption+~\cite{AffectGPT} & V,A,T & CMER & 31,327 & - & Model-led+Human-assisted & Multimodal Emotion Instruction \\
& MER-Caption~\cite{AffectGPT} & V,A,T & CMER & 115,595 & - & Model-led+Human-assisted & Multimodal Emotion Instruction \\
& AMT~\cite{EmoVerse} & I,V,T & FER,CMER & 32,940 & 10 & Human-led+Model-assisted & Affective Multitask \\
\midrule
% \multicolumn{7}{c}{\textit{\textbf{Evaluation Benchmark}}} \\
% \midrule
\multirow{12}{*}{\rotatebox{90}{Evaluation Benchmark}}
& VECBench~\cite{EmoCaliber} & I & GVEC & 8,235 & - & Human & Systematic Organization \\
& EEmo-Bench~\cite{EEmoBench} & I,T & GVEC & 6,733 & 7 & Human & Image-evoked Emotions \\
& MVEI~\cite{MVEI} & I,T & GVEC & 3,086 & OV & Model-led+Human-assisted & Emotion Statement Judgment \\
& MM-BigBench~\cite{MMBigBenchEM} & I,T & VTSA & 31,902 & - & Human & Systematic Organization \\
& FaceBench~\cite{FaceBench} & I & FER & 73,760 & 7 & Human-led+Model-assisted & Face Analysis \\ 
& EIBench~\cite{EIBench} & I,T & FER & 1,665 & 4 & Model-led+Human-assisted & Emotion Interpretation \\
& MTMEUR~\cite{MTMEUR} & V,T & CMER & 1,451 & 7 & Model-led+Human-assisted & Multi-turn Understanding \\
& Hi-EF~\cite{Hi-EF} & V,A,T & CMER & 3,069 & 7 & Human & Emotion Forecasting \\
& CA-MER~\cite{MoSEAR} & V,A,T & CMER & 1,500 & 9 & Model-led+Human-assisted & Emotion Conflicts \\
& MER-UniBench~\cite{AffectGPT} & V,A,T & CMER & 12,799 & - & Human & Systematic Organization \\
& OV-MERD~\cite{OVMER} & V,A,T & CMER & 332 & OV & Human-led+Model-assisted & Open-vocabulary \\
& EmotionHallucer~\cite{EmotionHallucer} & I,V,A,T & CMER & 2,742 & - & Human-led+Model-assisted & Emotion Hallucinations \\
\bottomrule
\end{tabular}
}
\caption{Emotion Data Annotation methods with their most critical and distinguishing attributes. We consider a dataset comprises modality ``T'' if the text carries information beyond basic task description. ``-" indicates descriptive/compound dataset. ``OV" denotes open-vocabulary.}%, which are categorized into training datasets and evaluation benchmarks. Important attributes such as modality, availability, sample size, label count, annotator types, and target tasks are listed. “I”, “A”, “V”, and “T” stand for image, audio, video, and text, respectively.
\label{tab:dataset}
\end{table*}
\begin{figure}[tb]
    \centering
    \includegraphics[width=1\linewidth]{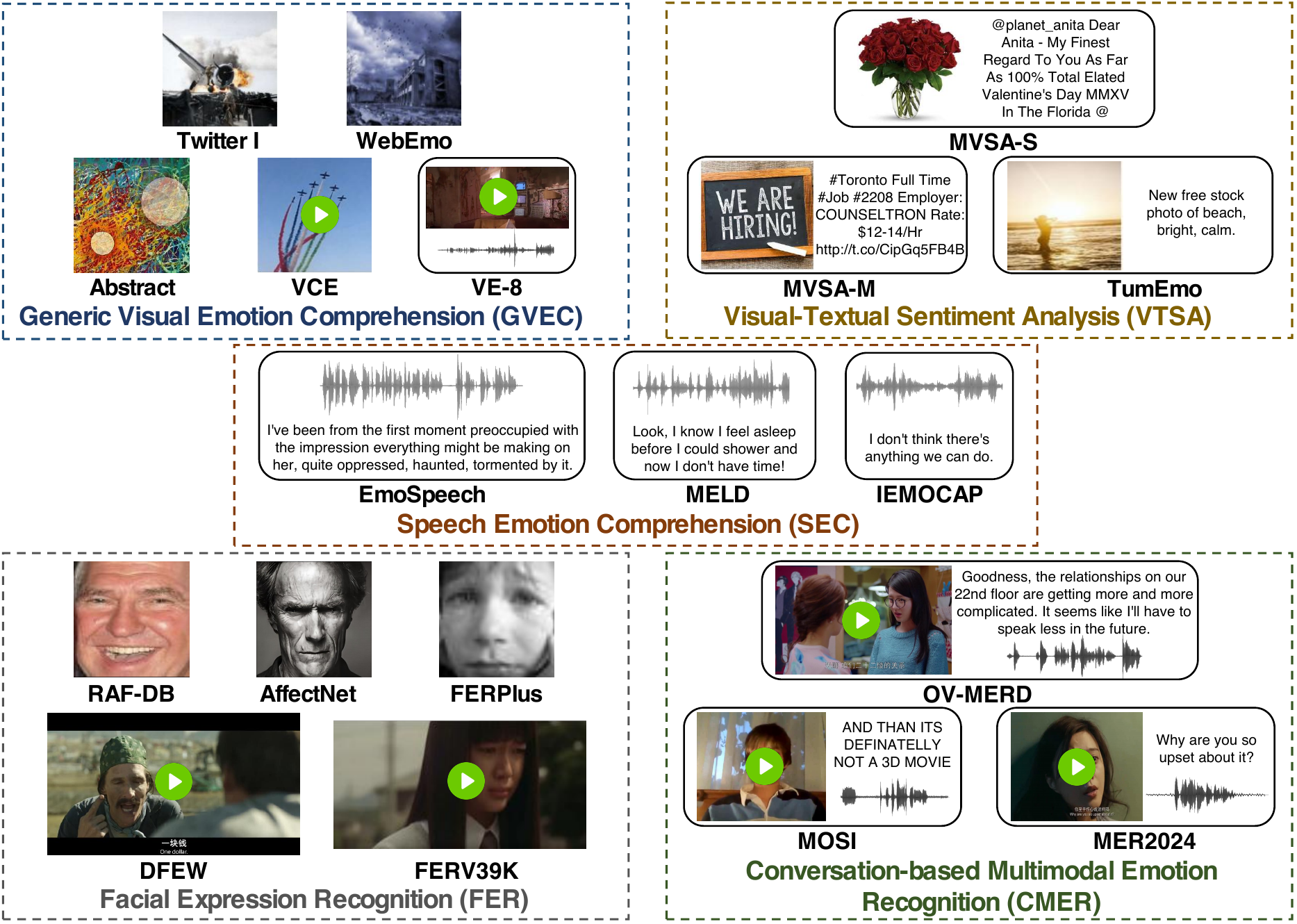}
    \caption{Visualization of samples from representative datasets of the five mainstream MER sub-tasks.}
    \label{fig:problem}
\end{figure}
%Accurately discerning emotional information from multimodal media remains a formidable challenge for MLLMs, primarily due to the inherent complexity of emotion data and the limitations of current model architectures. In this section, we define key MLLM tasks in emotion recognition and highlight three critical challenges that hinder progress in this field, each corresponding to a fundamental gap in affective computing research and application. 
% Before taxonomizing methods under the MER-with-LLMs paradigm, we first provide a formal task definition. 
Compared with traditional MER \cite{survey}, we extend the MER-with-LLMs paradigm to a broader setting. Under this paradigm, LLMs serve as a text-centered orchestrator that processes affective information from multimodal inputs containing at least one non-textual modality. 

Based on this definition, we first provide a symbolic formulation before taxonomizing existing methods.
%Compared with traditional MER, the MER-with-LLMs paradigm extends emotion understanding to broader tasks, where LLMs serve as language-centered reasoning modules for processing affective information from modalities beyond pure text inputs. To better characterize this paradigm, we first provide a formal task definition and clarify its boundary with related paradigms before taxonomizing existing methods.
% In general, a multimodal sample $X$ may consist of four modalities: $I$ (image), $V$ (video), $A$ (audio), and $T$ (text). 
In general, a multimodal sample $X$ can consist of various modalities, such as $I$ (image), $V$ (video), $A$ (audio), $T$ (text), or even physiological signals like electroencephalogram (EEG). Although the latter holds critical application value in certain areas~\cite{EEG1,MIR-EEG}, this paper primarily focuses on $I$, $V$, $A$, and $T$ modalities, considering their widespread availability.
Together with a text instruction $Q$ that specifies the task requirements, the sample is fed into an MLLM $\pi_\theta$. To process these inputs, the MLLM typically employs modality-specific encoders and adapters to transform $I$, $V$, and $A$ into multimodal embeddings, which are then combined with the tokenized $T$ and $Q$ to generate a textual response $R=[r_1,r_2,\cdots,r_n]$. Each token is generated by maximizing the conditional likelihood given the multimodal inputs and the preceding responses:
\begin{equation}
    r_i = \underset{r \in \mathcal{V}}{\arg\max} \, \pi_\theta(r|X, Q, R_{<i}),
\end{equation}
Here, $\mathcal{V}$ denotes the vocabulary of the MLLM. Depending on downstream requirements, the desired responses can take the form of emotion categories or tailored explanations. Separated by input domains and modalities, we divide MER into five mainstream sub-tasks, with samples visualized in Fig.~\ref{fig:problem}:
%Formally, for each sample \( X \), its content can potentially include multiple modalities, such as video, image, audio, and text. We denote these modalities as \( I \) (image), \( A \) (audio), \( V \) (video), and \( T \) (text), and we will use these abbreviations throughout the paper. Given an instruction \( Q \), the goal is to generate the correct response \( R \). For visual input \( X_V \), the MLLM uses a vision encoder to map it to a latent space \( Z_V \), and applies an adapter \( G_V \) to obtain visual tokens \( H_V = G_V(Z_V) \). Similar processing is done for audio and image inputs. After obtaining these tokens, they are concatenated and passed to the LLM decoder. The objective is to maximize the likelihood of the response \( R \), conditioned on the multimodal content \( (X_V, X_I, X_A, X_T) \) and the user instruction \( Q \). We represent the responses as \( R = \{r_{i}\}_{i=1}^{L_{r}}\), where \( L_{r} \) is the number of tokens:
% \begin{equation*}
% P(R \mid X, Q) = \prod_{l=1}^{Lr} P(r_{l} \mid X, Q, R_{<l}).
% \end{equation*}
%In this MLLM autoregressive equation, \( r_{l} \) is the current token to be predicted, and \( R_{<l} = \{r_{i}\}_{i=1}^{L_{r}} \) is the previously generated tokens, which serve as additional conditioning during training. Based on the emotion labels \( L_{emo} \) in the output response \( R \), emotion recognition can be classified into several categories:
\begin{itemize}
  \item \textit{Generic Visual Emotion Comprehension (GVEC)} aims to predict viewer-side emotional responses elicited by visual stimuli. $X=\{I\}/\{V\}/\{V,A\}$ consists of images or videos from arbitrary domains, such as artworks, landscapes, movies, or human-shared content.
  \item \textit{Visual-Textual Sentiment Analysis (VTSA)} aims to identify the emotion embedded in image–text pairs. $X=\{I, T\}$ is typically image–text posts collected from social media platforms such as Twitter and Tumblr.
  \item \textit{Speech Emotion Comprehension (SEC)} aims to describe emotions conveyed by speeches, commonly through classification or captioning. $X=\{A\}$ is a speech segment extracted from TV shows or purposely recorded.
  \item \textit{Facial Expression Recognition (FER)} aims to infer human emotional states from facial expressions. $X=\{I\}/\{V\}$ focuses on facial cues, typically in the form of a face-centered static image or dynamic video.
  \item \textit{Conversation-based Multimodal Emotion Recognition (CMER)} extends FER to a conversational setting, aiming to analyze a speaker’s emotional state from audiovisual dialogues. $X=\{V, A,  T\}/\{V, T\}$ is typically sourced from movies or TV shows.
\end{itemize}
% Although the inputs for GVEC and FER tasks are inherently unimodal, the language-centric nature requires MLLMs to always process text instructions, which expands them to a seamless multimodal setting. We therefore discuss them together.

\begin{figure}[tb]
    \centering
    \includegraphics[width=1\linewidth]{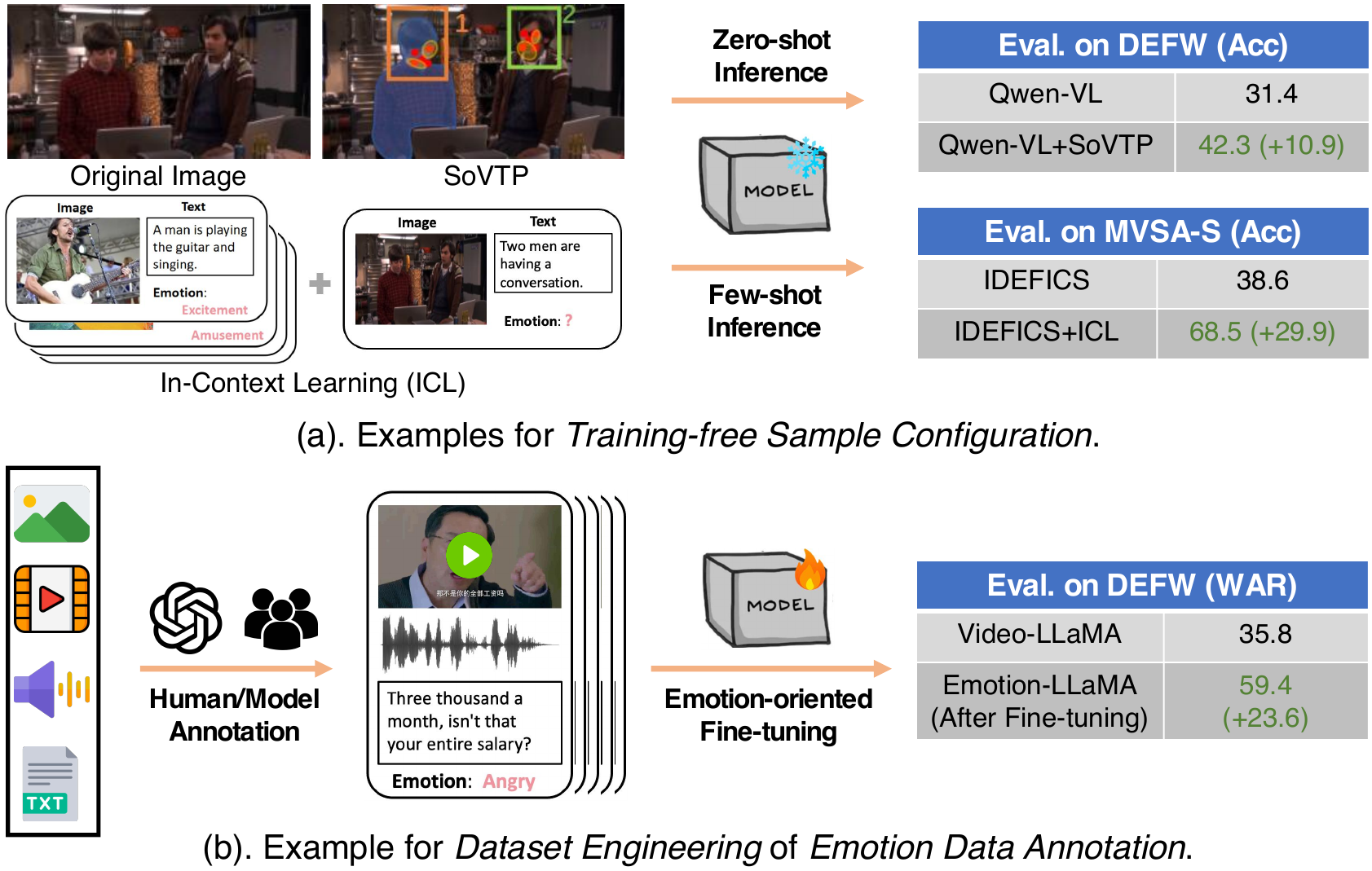}
    \caption{Illustration of Affective Data Augmentation methods.}
    \label{fig:augmentation}
\end{figure}
\begin{table*}[tb]
\centering
\resizebox{0.92\textwidth}{!}{
\begin{tabular}{@{}lcccccc@{}}
\toprule
\multirow{2}{*}{\textbf{Method}} & \multirow{2}{*}{\textbf{Modality}} & \multirow{2}{*}{\textbf{Sub-task}} & \multirow{2}{*}{\textbf{Technique}} & \multicolumn{2}{c}{\textbf{Encoder Adaptation}} & \multirow{2}{*}{\textbf{Key Design}} \\ 
\cmidrule(lr){5-6} & & & & Image/Video & Audio & \\
\toprule
SEPM~\cite{SEPM} & I & GVEC & Zero-Shot & Sharpened & - & Coarse to Fine Inference \\
MERMAID~\cite{MERMAID} & I & GVEC & Zero-Shot & Augmented  & - & Self-reflective Agent \\
StimuVAR~\cite{StimuVAR} & V & GVEC & SFT & Sharpened & - & Recognition before Reasoning \\
EmoDETective~\cite{EmoDETective} & V,A & GVEC & SFT+RL  & Basic & Basic & Fact before Emotion \\
ExpLLM~\cite{ExpLLM} & I & FER & SFT & Basic  & - & Multi-turn Conversation \\
EmoLA~\cite{EmoLA} & I & FER & SFT & Augmented  & - & Dual-perspective Encoding \\
FaceInsight~\cite{FaceInsight} & I & FER & SFT & Augmented  & - & Correlation\&Logical Constraint \\
BLSP-Emo~\cite{BLSP-Emo} & A & SEC & SFT & - & Basic & Unimodal before Multimodal \\
SECap~\cite{SECap} & A & SEC & SFT  & - & Sharpened & Q-former-based Compression \\
EmoVerse~\cite{EmoVerse} & V,T & CMER & SFT & Basic & - & Multi-task Learning \\
Emotion-LLaMA~\cite{Emotion-LLaMA} & V,A,T & FER, CMER & SFT & Augmented & Basic & Multi-view Encoding \\
% EmoChat~\cite{EmoChat} & V,A,T & GVEC,VTSA,CMER & SFT  & Basic+Expert & Basic & Feature Alignment \\
\multirow{2}{*}[0.2cm]{EmoChat~\cite{EmoChat}} & I,V,A,T & \makecell{GVEC, VTSA,\\FER, CMER} & SFT & Augmented & Basic & Feature Alignment \\
\bottomrule
\end{tabular}
}
\caption{Perceptual Emotion Mapping methods with their key attributes. ``Encoder Adaptation'' characterizes how each method adapts the encoding process beyond parameter updating. ``Basic'' indicates no adaptation. SFT: Supervised Fine-tuning. RL: Reinforcement Learning. }
\label{tab:perception}
\end{table*}
%Models proposed for Perceptual Emotion Mapping. They are distinguished based on their modality, finetuned components, encoding perspectives (visual, facial, and acoustic), and additional design elements. Ecoding perspectives characterize how each method encode multimodal information, where 'Basic' indicates a common encoder.
%
% Existing MLLMs primarily address three fundamental challenges to enhance their emotion recognition capabilities: affective data scarcity, multimodal affective gap, and affective interpretation opacity. In this paper, we provides a comprehensive taxonomy of methods targeting these challenges in MER with LLMs. \\

% Although SEC, FER, and partial GVEC take unimodal inputs from definition, text instructions are indispensable for MLLMs, causing these tasks to be handled in a manner essentially indistinguishable from multimodal settings. As a result, we also include them under the MER-with-LLMs paradigm. 
On top of the above formulation, we present the taxonomy of MER-with-LLMs in Fig.~\ref{fig:taxonomy}. We categorize existing methods into three branches according to the challenges they focus on and provide a high-level overview below: \\
% While GVEC and FER tasks
% Existing MLLMs tackle three primary challenges to improve emotion recognition: the scarcity of affective data, the multimodal affective gap, and the opacity in affective interpretation. This paper presents a comprehensive taxonomy of methods designed to address these challenges within the context of MER using LLMs. % innovative strategies such as few-shot and zero-shot learning approaches,  the development of specialized datasets and benchmark infrastructures for affective recognition
\textbf{Affective Data Augmentation.} Existing methods for handling Affective Data Scarcity are divided into two branches. \textit{Training-free Sample Configuration} aims to maximize the utilization of available data, typically by augmenting the input context while freezing MLLM parameters. \textit{Emotion Data Annotation} focuses on augmenting the volume of data, either by engineering training data for fine-tuning or by constructing benchmarks that enable more emotion-aware evaluation. \\
%tackles the issue of limited emotional data for training emotion prediction models. This challenge is mitigated through \textit{Training-free Sample Configuration for Emotion Inference}, which enable models to generalize to new tasks with minimal annotated data. Additionally, \textit{Affective Recognition Downstream Task Infrastructure} highlights the need for specialized datasets and benchmark infrastructures, essential for improving model performance in downstream emotion recognition tasks. These efforts are crucial for improving performance in practical applications, where large-scale labeled datasets are often limited, thus facilitating the generalization of MLLMs to diverse real-world scenarios. % (data input stage)
\textbf{Multimodal Affective Representation.} To handle Multimodal Affective Gap, a line of research focuses on feeding multimodal embeddings with richer affective representations into LLMs. According to their primary emphasis, we further divide these methods into two branches: \textit{Perceptual Emotion Mapping}, which refines the perception of intra-modality affective cues by optimizing the encoding process, and \textit{Multimodal Emotion Coordination}, which enhances the modeling of inter-modality coordination through comprehensive fusion of affective cues from multiple sources. \\
%focuses on enhancing the model's ability to perceive and represent nuanced emotional cues across multiple modalities. Key techniques in this stage include \textit{Perceptual Emotion Mapping}, which refines the model’s sensitivity to subtle emotional cues by optimizing encoder architectures, and \textit{Multimodal Emotion Coordination}, which improves the model's ability to integrate and fuse emotional information from different sources. These techniques ensure that MLLMs can generate a more coherent and accurate emotional representation by addressing the complexity and potential conflicts of multimodal inputs. % (emotion modeling stage)
\textbf{Multimodal Affective Reasoning.} To address Affective Interpretation Opacity, some studies encourage MLLMs to generate step-by-step reasoning before producing final emotion predictions, thereby enhancing explainability. Pioneering works primarily focus on the accuracy and reliability of such reasoning, which we categorize as \textit{Emotion Explanation and Hallucination}. More recent studies further explore accommodating the inherent subjectivity of emotion perception, which we classify as \textit{Subjective Emotion Reasoning}. 

Below, we elaborate on each branch. Methods that fall into overlapping regions are classified by their primary focus.
%is concerned with improving the interpretability and reasoning capabilities of emotion models. The goal is to enable MLLMs to provide transparent, understandable, and reliable emotional inferences. Key advancements in this domain include \textit{Emotion Explanation and Hallucination}, which offers human-readable justifications for model predictions, and strategies to minimize hallucinations during emotional inference. Additionally, \textit{Subjective Emotion Reasoning} enables the model to dynamically reason about emotions in context. By addressing the opacity in emotion inference, this stage is vital for ensuring that MLLMs are capable of making emotion-based decisions that are both interpretable and robust across diverse applications, such as healthcare, education, and emotional support systems. % (output stage)
% Overview of Taxonomy
% \section{Taxonomy of MER in MLLMs} \label{sec:taxonomy}
% \input{sections/taxonomy}
% Methodology
\section{Affective Data Augmentation} \label{sec:augmentation}
% Training datasets and evaluation benchmarks are crucial for advancing and assessing the emotion recognition capabilities of MLLMs. To address the shortage of emotional data, researchers have introduced Affective Data Augmentation through two key approaches: Training-Free Sample Configuration for Emotion Inference and Affective Recognition Downstream Task Infrastructure, both aimed at improving model performance in MER. \\
\textbf{Training-free Sample Configuration.} 
% Rather than relying solely on text-based sample configurations or prompts~\cite{ECR-Chain,NegativePrompt}, this line of work emphasizes maximizing the utility of existing multimodal samples to handle Affective Data Scarcity.
To handle Affective Data Scarcity, this line of work emphasizes maximizing the utility of existing samples.
This is achieved by carefully configuring the samples provided as input to MLLMs, a process that introduces human-prior-inspired biases or demonstrations. Such strategies are cost-efficient and are particularly useful when additional data is unavailable.
%In the absence of emotional data, the model's emotion recognition ability can be enhanced by adjusting the sample configuration during reasoning. This approach improves performance without requiring large annotated datasets or significant computational resources, making it particularly useful when training data is limited or unavailable.

Under zero-shot settings, the biases are injected through extracting key image features, allowing MLLMs to focus on regions that convey the most salient emotional cues. For example, initial explorations~\cite{EMOTIC} utilize CLIP~\cite{CLIP} to extract emotional information from bounding boxes around individuals in images, generating captions that reflect emotional content. While bounding boxes alone provide useful spatial localization, they often lack the granularity needed for nuanced MER. To overcome this limitation, several approaches~\cite{SoV,SoVTP} incorporate multi-level visual cues, including spatial localization, facial geometry, and action units, enabling a more comprehensive analysis of human emotional information.
% For instance, NarraCap~\cite{EMOTIC} utilizes CLIP to extract emotional information from bounding boxes around individuals in images, generating captions that reflect emotional content. While bounding boxes alone offer useful spatial localization, they often lack the granularity needed for nuanced emotion recognition. To address this limitation, SoV~\cite{SoV} enhances the approach by progressively incorporating facial bounding boxes, numbered boxes to differentiate faces, and facial landmarks to analyze spatial relationships, thereby improving the accuracy of MER. Building on SoV, SoVTP~\cite{SoVTP} further extends this by adding body masks for person identification and tracking, and incorporates facial action units to provide a more detailed analysis of facial muscle movements.
Inspired by the success of In-Context Learning (ICL), Wu \textit{et al.}~\shortcite{ICL} unleash the emotion perception of MLLMs in a few-shot setting. By optimizing the retrieval, presentation, and distribution of demonstrations, this method allows MLLMs to generalize effectively from minimal labeled data, further improving emotion inference in multimodal contexts. We illustrate them in Fig.~\ref{fig:augmentation}(a).
% In the few-shot setting, particularly within the framework of In-Context Learning (ICL)~\cite{ICL}, the model's performance in Multimodal Sentiment Analysis (MSA) can be significantly enhanced by optimizing the retrieval, presentation, and distribution of demonstrations. By carefully selecting and arranging a limited set of sample configurations, this method allows the model to generalize more effectively from minimal labeled data, further improving emotion inference in multimodal contexts.

% 
\begin{table*}[tb]
\centering
% \footnotesize
% \resizebox{\dimexpr\textwidth/2}{!}{
% \resizebox{\columnwidth}{!}{
\resizebox{\textwidth}{!}{
\begin{tabular}{@{}lccccc@{}}
\toprule
\textbf{Method} & \textbf{Modality} & \textbf{Sub-task}  & \textbf{Technique} & \textbf{Encoder} & \textbf{Abbreviated Fusion Mechanism}  \\
\toprule
EmoVIT~\cite{EmoVIT} & I & GVEC  & SFT & I &Q-former(q=[Learned,T],k\&v=I)\\ 
Omni-SILA~\cite{Omni-SILA} & V,T & GVEC  & SFT & V,$\text{V}_\text{f}$,$\text{V}_\text{h}$,$\text{V}_\text{o}$ & Mixture-of-Experts(V,$\text{V}_\text{f}$,$\text{V}_\text{h}$,$\text{V}_\text{o}$) \\
% EMO-LLaMA~\cite{EMO-LLaMA} & V,T & Vicuna & Adapter, LLM & Face-Info-Mining(F,V)+Clue(T,F,V)+F+T & Dataset Engineering \\
FEALLM~\cite{FEALLM} & I & FER  & SFT & I,$\text{I}_\text{f}$ & Cross-attention(q=I,k\&v=$\text{I}_\text{f}$)  \\
Omni-Emotion~\cite{Omni-Emotion} & V,A,T & FER, CMER  & SFT & V,$\text{V}_\text{f}$,A  & [V,$\text{V}_\text{f}$] \\
AffectGPT~\cite{AffectGPT} & V,A,T & CMER  & SFT & V,A & Self-attention([V,A])  \\
M$^{3}$F~\cite{M3F} & V,A,T & CMER  & SFT & V,A & Q-former(q=Learned,k\&v=V)+Q-former(q=Learned,k\&v=A) \\
% \multirow{2}{*}[0.2cm]{EMO-LLaMA~\cite{EMO-LLaMA}} & I,V,T & FER, CMER  & SFT & \makecell{Face-Info-Mining(F,V)+\\Clue(T,F,V)+F+T} \\
EMO-LLaMA~\cite{EMO-LLaMA} & I,V,T & FER, CMER & SFT & V,$\text{V}_\text{f}$ & Q-former(q=[Learned,T],k\&v=[V,$\text{V}_\text{f}$])+Self-attention([V,$\text{V}_\text{f}$]) \\
\bottomrule
\end{tabular}
}
\caption{Multimodal Emotion Coordination methods with their key attributes. ``Encoder'' lists the adopted encoders, where $\text{I}_\text{f}$ represents image facial encoder, $\text{V}_\text{f}$, $\text{V}_\text{h}$, and $\text{V}_\text{o}$ denote video facial, human, object encoders, respectively. Operation [$\cdot$] represents concatenation.}
\label{tab:coordination}
\end{table*}
%Comparison of different Multimodal Emotion Coordination methods. They integrate multiple modalities for emotion recognition, highlighting the base models, finetuned components, and fusion mechanisms used to combine these modalities effectively. F, HA, and OR denote facial expression, human action, and object relation, respectively.
\noindent
\textbf{Emotion Data Annotation.} When additional data is permitted, annotating the desired emotion-oriented datasets serves as a more straightforward approach. In this line of work, we separately introduce the training datasets and evaluation benchmarks, both of which are summarized in Tab.~\ref{tab:dataset}.
%Even though sample configuration improves MLLM's ability on MER downstream tasks, it is the construction of training datasets and the evaluation benchmarks that can truly enhance the model's emotion recognition capabilities, as summarized in in Table \ref{tab:dataset}.

\noindent
\textbf{Dataset Engineering.} 
This branch enhances the intrinsic capabilities of MLLMs by constructing training datasets, as illustrated in Fig.~\ref{fig:augmentation}(b). For the GVEC task, EmoVIT~\cite{EmoVIT} leverages advanced proprietary models to annotate the first emotion-centric visual instruction tuning dataset, and VEC-CoT~\cite{EmoCaliber} subsequently contributes a larger-scale structured affective reasoning dataset. For the FER task, FABA-Instruct~\cite{EmoLA} considers both action units and emotions, introducing the first dataset of its kind. AMT~\cite{EmoVerse} further extends to a multi-task scenario. The CMER task has received the most attention, likely due to its highly complex input modalities, with different methods exploring complementary scenarios. Specifically, DEEMO~\cite{DEEMO} is designed to preserve privacy; EmoCause~\cite{EmoDETective} and MEC$^{4}$~\cite{M3F} focus on understanding emotion cause under multi-attribute and multilingual settings; MESC~\cite{MESC} targets therapeutic and counseling scenarios; and MERR~\cite{Emotion-LLaMA} and MER-Caption~\cite{AffectGPT} address regular CMER tasks, characterized by their high quality and large scale. These datasets provide essential prerequisites for subsequent emotion-oriented optimization.

\noindent
\textbf{Benchmark Construction.} 
Beyond training, evaluation also plays an indispensable role in the development of MER-with-LLMs, whose evolution exhibits unexpected consistency across MER sub-tasks. Pioneering efforts focus on systematically integrating existing datasets, such as VECBench~\cite{EmoCaliber}, MM-BigBench~\cite{MMBigBenchEM}, and MER-UniBench~\cite{AffectGPT}. Subsequent works aim to shed light on deeper insights. For example, EEmo-Bench~\cite{EEmoBench} unifies evaluation of perception, description, ranking, and assessment; FaceBench~\cite{FaceBench} encompasses perception for fine-grained facial details; EIBench~\cite{EIBench} delves into both explicit and implicit factors that drive emotional responses; Hi-EF~\cite{Hi-EF} proposes forecasting future emotional trajectories; and MTMEUR~\cite{MTMEUR} probes MLLMs’ understanding of multi-turn emotional dialogues.

More recent efforts push further toward deeper emotional dynamics. OV-MERD~\cite{OVMER} annotates samples with multi-label, open-vocabulary emotion descriptions, and MVEI~\cite{MVEI} introduces the emotion statement judgment task, both of which incorporate the previously overlooked issue of subjectivity in evaluation. CA-MER~\cite{MoSEAR} examines how models handle emotional conflicts across modalities, which is a practical yet underexplored problem. EmotionHallucer~\cite{EmotionHallucer} conducts an in-depth investigation of hallucinations in emotional reasoning, advancing reliability. Collectively, these datasets and benchmarks act as guiding beacons for the MER-with-LLMs paradigm, providing anchors and directions.

\begin{figure}[t]
    \centering
    \includegraphics[width=1\linewidth]{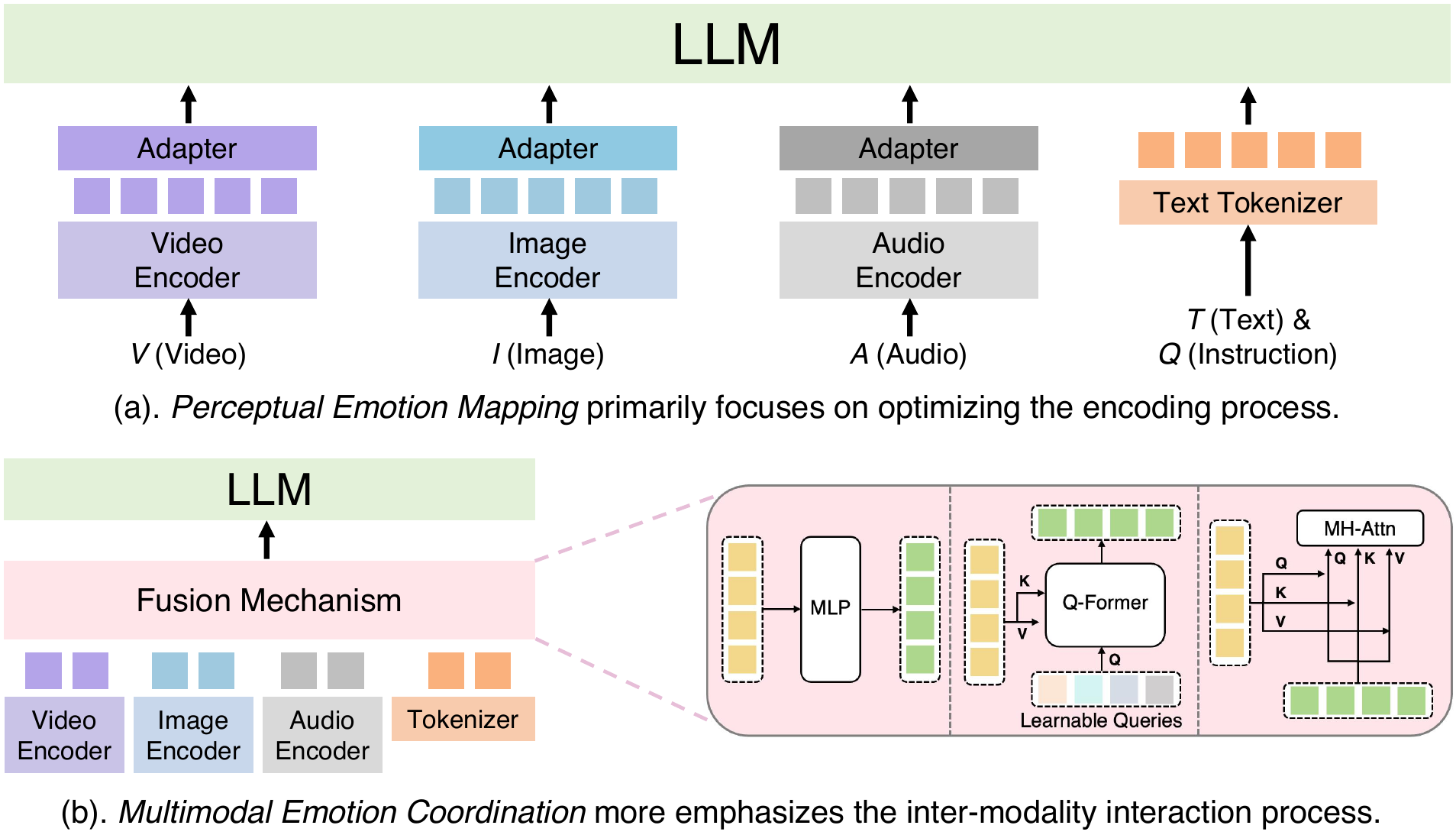}
    \caption{Differences in methodological focuses of the two types of Multimodal Affective Representation methods.}
    \label{fig:representation}
\end{figure}
\begin{table*}[tb]
\centering
\resizebox{0.9\textwidth}{!}{
\begin{tabular}{@{}clccccc@{}}
\toprule
& \textbf{Method} & \textbf{Modality} & \textbf{Sub-task}  & \textbf{Technique} & \textbf{RL Reward} & \textbf{Key Focus}  \\ 
\toprule
\multirow{8}{*}{\rotatebox{90}{Explanation}}
& MulCoT-RD~\cite{MulCoT-RD} & I,T & VTSA & SFT & - & Computational Efficiency  \\
& Facial-R1~\cite{Facial-R1} & I & FER & SFT+RL & Rule & Explanation Consistency  \\
& AlignCap~\cite{AlignCap} & A & SEC & SFT+RL & LLM-as-judge & Hallucination Mitigation \\
& Affective-CoT~\cite{Affective-CoT} & V,A,T & CMER & Few-shot & - & Emotion Conflict Resolving  \\
& PEP-MEK~\cite{EmotionHallucer} & I,V,A,T & GVEC, CMER &  Zero-shot & - & Hallucination Mitigation  \\
& ERV~\cite{ERV} & V,A,T & FER, CMER & SFT+RL & Model & Explanation Consistency  \\
%& MoSEAR~\cite{MoSEAR} & V,A,T & FER,CMER & SFT & - & Emotion Conflict Resolving  \\
& FIRES~\cite{FIRES} & V,T & $\text{MESC}^*$ & SFT+RL & Rule & Thinking Flexibility  \\
& MultiMood~\cite{MultiMood} & V,A,T & CMER, $\text{MESC}^*$ & SFT+RL & LLM-as-judge & Reinforce Trustworthiness  \\ 
\midrule
\multirow{3}{*}{\rotatebox{90}{Subject.}}
& EmoCaliber~\cite{EmoCaliber} & I & GVEC & SFT+RL & Rule & Confidence Verbalization  \\
& Agent-MER~\cite{Agent-MER} & V,A,T & CMER & Zero-shot & - & Open-vocabulary  \\
& AffectGPT-R1~\cite{AffectGPT-R1} & V,A,T & CMER & SFT+RL & Rule & Open-vocabulary  \\
\bottomrule
\end{tabular}
}
\caption{Multimodal Affective Reasoning methods with their key attributes. $\text{MESC}^*$ stands for Multimodal Emotional Support Conversation, which takes similar inputs as CMER but focuses more on generating supportive dialogue to alleviate stress for the user.}
\label{tab:Understanding}
\end{table*}
% Comparison of different Multimodal Affective Reasoning methods. The upper section covers approaches to emotion explanation and hallucination mitigation, while the lower section focuses on subjective emotion reasoning. They are characterized by modality, base models, training strategies, finetuned components, target affective reasoning tasks, and reward source. The `Reward Source' denotes the supervision signal used for affective reasoning enhancement, including LLM, explicit rule-based functions (Rule), and learned verifier or discriminator models (Model).
\section{Multimodal Affective Representation} \label{sec:representation}
% Even with sufficient training datasets and evaluation benchmarks, further improving a model’s ability to recognize emotional cues from characters in audio and video, as well as to enhance its perception of multimodal emotional representations, remains a critical challenge. Owing to the complexity and entanglement of factual and affective features, general MLLMs often rely on superficial interpretations of emotional signals and fail to capture fine-grained emotional cues, which in turn leads to emotion hallucinations and errors in emotional reasoning.
As illustrated in Fig.~\ref{fig:representation}, to tackle Multimodal Affective Gap, current methods fall into two categories: one bridges intra-modality gaps by refining the encoding process, while the other handles inter-modality gaps by modeling interactions.

\noindent
\textbf{Perceptual Emotion Mapping.} 
Multimodal encoders are typically pre-trained on general-purpose data, which makes them less effective when directly applied to capturing affective cues. Existing solutions are largely shared across different subtasks. An intuitive one is curating data and arranging training schedules, thereby adjusting the parameters of encoders or adapters to a more emotion-aligned direction. For instance, EmoDETective~\cite{EmoDETective} introduces a progressive training strategy that transitions from factual detection to emotional conclusion; ExpLLM~\cite{ExpLLM} constructs an Exp-CoT engine to synthesize data and improves FER through multi-turn conversations; BLSP-Emo~\cite{BLSP-Emo} first teaches models to understand emotional transcripts before incorporating audio signals; and EmoVerse~\cite{EmoVerse} jointly optimizes multiple tasks and ensures smooth training by controlling the task ratios.

In parallel, directly modifying the encoded embeddings through sharpening or augmentation serves as another established solution. Sharpening promotes the capture of affective cues by filtering out redundant features before the LLM. For instance, SEPM~\cite{SEPM} applies coarse-to-fine inference to remove emotion-irrelevant visual tokens; StimuVAR~\cite{StimuVAR} proposes event-driven frame sampling and emotion-triggered tube selection strategies; and SECap~\cite{SECap} employs a learnable Q-former to compress audio embeddings. By contrast, augmentation complements affective cues by introducing additional features: MERMAID~\cite{MERMAID} treats MLLMs as agents and augments the original image through self-reflection; EmoLA~\cite{EmoLA}, Emotion-LLaMA~\cite{Emotion-LLaMA}, FaceInsight~\cite{FaceInsight}, and EmoChat~\cite{EmoChat} implement augmentation more straightforwardly by incorporating expert encoders, with the latter two further introducing consistency constraints to facilitate feature alignment. These methods are collectively organized in Tab.~\ref{tab:perception}.

\noindent
\textbf{Multimodal Emotion Coordination.} To model the synergy of inter-modality affective cues, current works rely on modality fusion to enable thorough interactions before feeding the representations into the LLM. We summarize these methods in Tab.~\ref{tab:coordination}. Overall, they primarily adopt concatenation, Q-former, and attention-based mechanisms, often accompanied by the encoder augmentation strategies discussed earlier.

Specifically, Omni-Emotion~\cite{Omni-Emotion} finds that video-level concatenation outperforms frame-level interactions. FEALLM~\cite{FEALLM} and AffectGPT~\cite{AffectGPT} employ vanilla attention mechanisms, where the former focuses on fusion across different hierarchical levels and scales, while the latter emphasizes holistic interaction. EmoVIT~\cite{EmoVIT} and M$^{3}$F~\cite{M3F} favor Q-former-based designs, leveraging learnable queries to extract instruction-aligned cues or retrieve memory fragments. EMO-LLaMA~\cite{EMO-LLaMA} combines attention with Q-former mechanisms to inject rich facial knowledge into image embeddings. Notably, Omni-SILA~\cite{Omni-SILA} adopts an alternative fusion paradigm based on Mixture-of-Experts for effective multi-source integration.
% , which has also been empirically verified to be effective for integrating multi-source information.
In general, these diversified successes highlight the inherent complexity and future potential of bridging inter-modality affective gaps.

\begin{table*}[tb]
\centering
\resizebox{\textwidth}{!}{
\begin{tabular}{@{}cllclcr@{}}
\toprule
 & \textbf{Method} & \textbf{Taxonomy} & \textbf{Technique} & \textbf{Dataset} & \textbf{Metrics} & \textbf{Results} \\ 
\toprule
% \multicolumn{6}{c}{\textit{\textbf{Basic Emotion Recognition}}} \\
% \midrule % Add MERMIAD, 
\multirow{3}{*}{\rotatebox{90}{\makecell{GVEC}}}
& SEPM~\cite{SEPM} & Perceptual Emotion Mapping & Zero-shot & EmoSet, Emotion6 & Acc $\uparrow$ & 56.24\%, 54.21\% \\
& MERMAID~\cite{MERMAID} & Perceptual Emotion Mapping & Few-shot & EmoSet, Emotion6 & Acc $\uparrow$ & 63.70\%, 61.90\% \\
& EmoVIT~\cite{EmoVIT} & Multimodal Emotion Coordination & SFT & EmoSet, Emotion6 & Acc $\uparrow$ & 83.36\%, 57.81\% \\
\midrule
% \multicolumn{6}{c}{\textit{\textbf{Sentiment Analysis}}} \\
\multirow{3}{*}{\rotatebox{90}{\makecell{VTSA}}}
% & GPT-4V~\cite{GPT-4V} & - & Zero-shot & MVSA-M & Acc & 66.82\% \\
& ICL~\cite{ICL} & Training-free Sample Configuration & Few-shot & MVSA-M & Acc $\uparrow$ & 69.50\% \\
& EmoChat~\cite{EmoChat} & Perceptual Emotion Mapping & SFT & MVSA-M & Acc $\uparrow$ & 72.70\% \\
& MulCoT-RD~\cite{MulCoT-RD} & Emotion Explanation and Hallucination & SFT & MVSA-M & Acc $\uparrow$ & 77.20\% \\
\midrule
\multirow{2}{*}{\rotatebox{90}{\makecell{SEC}}}
& SECap~\cite{SECap} & Perceptual Emotion Mapping & SFT & NNIME, EMOSEC & BLEU@4 $\uparrow$ & 5.8, 7.4 \\
& AlignCap~\cite{AlignCap} & Emotion Explanation and Hallucination & RL & NNIME, EMOSEC & BLEU@4 $\uparrow$ & 7.7, 9.8 \\
\midrule
% \multicolumn{6}{c}{\textit{\textbf{Facial Emotion Recognition}}} \\
% \midrule
\multirow{5}{*}{\rotatebox{90}{\makecell{FER}}}
& ExpLLM~\cite{ExpLLM} & Perceptual Emotion Mapping & SFT & RAF-DB & Acc $\uparrow$ & 91.03\% \\
& EmoLA~\cite{EmoLA} & Perceptual Emotion Mapping & SFT & RAF-DB & Acc $\uparrow$ & 92.05\% \\
& Facial-R1~\cite{Facial-R1} & Emotion Explanation and Hallucination & SFT+RL & RAF-DB & Acc $\uparrow$ & 92.10\% \\
& EMO-LLaMA~\cite{EMO-LLaMA} & Multimodal Emotion Coordination & SFT & DFEW & UAR $\uparrow$ & 60.23\% \\
% & Emotion-LLaMA~\cite{Emotion-LLaMA} & V,A,T & SFT & DFEW & UAR & 64.21\% \\
& ERV~\cite{ERV} & Emotion Explanation and Hallucination & SFT+RL & DFEW & UAR $\uparrow$ & 68.88\% \\
\midrule
% \multicolumn{6}{c}{\textit{\textbf{OV-MER}}} \\
% \midrule
\multirow{3}{*}{\rotatebox{90}{\makecell{CMER}}}
% & \multirow{2}{*}{\raisebox{1.5ex}{AffectGPT~\cite{AffectGPT}}} & SFT & \makecell{MER2023, MER2024, MELD, IEMOCAP,\\MOSI, MOSEI, SIMS, SIMS-v2, OV-MERD+} & WAF & \makecell{78.64\%, 78.80\%, 55.65\%, 60.54\%,\\ 81.30\%, 80.90\%, 88.49\%, 86.18\%, 62.52\%} \\
% & \multirow{2}{*}{\raisebox{1.5ex}{AffectGPT-R1~\cite{AffectGPT-R1}}} & SFT+RL & \makecell{MER2023, MER2024, MELD, IEMOCAP,\\MOSI, MOSEI, SIMS, SIMS-v2, OV-MERD+} & WAF & \makecell{84.51\%, 93.13\%, 66.71\%, 74.26\%,\\ 79.65\%, 80.18\%, 87.26\%, 85.75\%, 68.39\%} \\
& Emotion-LLaMA~\cite{Emotion-LLaMA} & Perceptual Emotion Mapping & SFT & MER2024, MOSI, OV-MERD+ & WAF $\uparrow$ & 73.62\%, 66.13\%, 52.97\% \\
& AffectGPT~\cite{AffectGPT} & Multimodal Emotion Coordination & SFT & MER2024, MOSI, OV-MERD+ & WAF $\uparrow$ & 78.80\%, 81.30\%, 62.52\% \\
& AffectGPT-R1~\cite{AffectGPT-R1} & Subjective Emotion Reasoning & SFT+RL & MER2024, MOSI, OV-MERD+ & WAF $\uparrow$ & 93.13\%, 79.65\%, 68.39\% \\
\bottomrule
\end{tabular}
}
\caption{Quantitative performance of representative methods of MER-with-LLMs. Results are fetched from the corresponding papers. In the ``Metrics'' column, ``Acc'' denotes accuracy, ``UAR'' denotes unweighted average recall, and ``WAF'' denotes average F-score.}
\label{tab:Results}
\end{table*}
% Illustration of reasoning
\begin{figure}[t]
    \centering
    \includegraphics[width=1\linewidth]{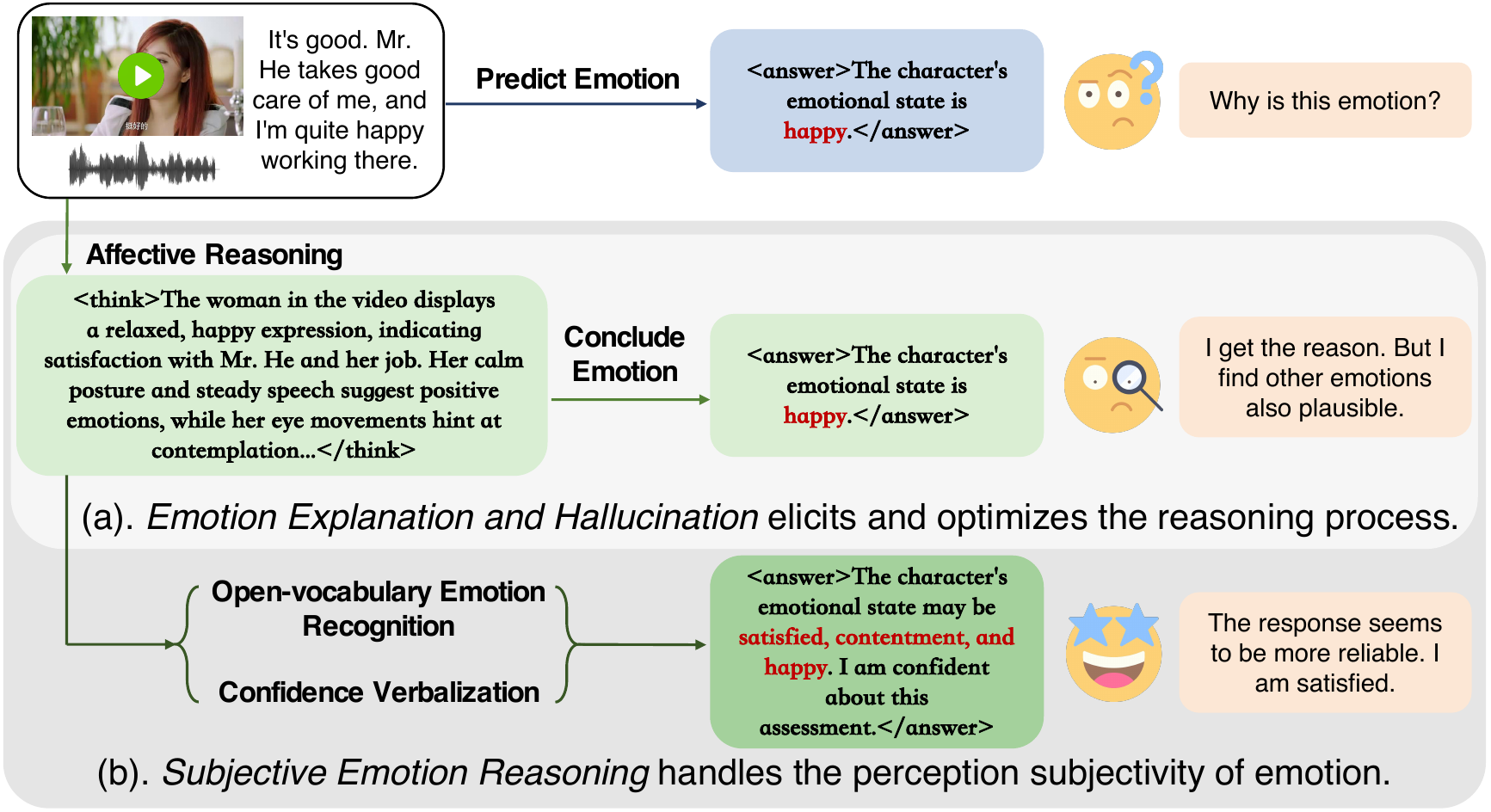}
    \caption{Evolution from sole emotion prediction, to explanation, and finally to subjectivity embracement within Multimodal Affective Reasoning methods.}
    \label{fig:reasoning}
\end{figure}
\section{Multimodal Affective Reasoning} \label{sec:reasoning}
%Once the model is capable of perceiving emotional data with fine granularity, a new direction for research is to explore how the model can explain the underlying causes of emotions and provide subjective emotion reasoning, as summarized in Table~\ref{tab:Understanding}.
% This would improve user trust and enable more user-friendly outputs, such as open-vocabulary responses. \\
% Enhancing interpretability also makes emotional reasoning more transparent, reducing hallucinations and increasing the reliability of emotional inferences. \\
As illustrated in Fig.~\ref{fig:reasoning}, directly predicting emotions may lead to user confusion regarding model decisions, which has motivated the emergence and development of the explainable MER task. This line of research brings diverse benefits: it not only enhances user trust but also helps uncover potential model deficiencies and optimization pathways.

\noindent
\textbf{Emotion Explanation and Hallucination.} 
% After incorporating emotion labels reasoning, Explainable MER (EMER)~\cite{EMER} extends this by providing explanations, as shown in Fig.~\ref{fig:reasoning}(a). 
Methods in this branch primarily focus on the usability and reliability of explanations. To address potential inconsistencies between reasoning and predictions, Facial-R1~\cite{Facial-R1} proposes a data-centric solution, which is also supported by an RL reward that encourages factual observations. ERV~\cite{ERV} trains a lightweight auxiliary model specifically to verify whether the answers are faithful to the reasoning. To mitigate hallucinations, PEP-MEK~\cite{EmotionHallucer} introduces a predict–explain–predict inference pipeline, inserting an additional stage to extract modality-specific and emotion-related knowledge. AlignCap~\cite{AlignCap} incorporates guidance from teacher models and proprietary LLMs to enforce consistency and rationality. 

Some other critical directions have also been explored. Affective-CoT~\cite{Affective-CoT} targets potential emotional conflicts between modalities by decoupling perception and reasoning, and gradually resolving contradictions through a few-shot, cognition-driven workflow. MulCoT-RD~\cite{MulCoT-RD} improves computational efficiency through distilling knowledge from heavy MLLMs to lightweight ones. FIRES~\cite{FIRES} equips the model with a flexible thinking pattern that dynamically incorporates ``visual scene'', ``emotion'', ``situation'', and ``response strategy'' on demand. MultiMood~\cite{MultiMood} reinforces user trust in emotional support dialogue by formulating psychology-based criteria for guidance.

\noindent
\textbf{Subjective Emotion Reasoning.} 
The inherent subjectivity of emotion perception allows different observers to produce different responses to the same content, posing a long-standing challenge to the MER field. Leveraging the MER-with-LLMs paradigm, several meaningful attempts have been made.

Some methods reformulate pre-defined emotion sets into open-vocabulary ones and allow multiple labels per sample, enabling models to recognize as many plausible emotions as possible. 
% Along this direction, Agent-MER~\cite{Agent-MER} emulates the human cognitive circle by transforming single-step emotion prediction into a structured, knowledge-guided process, and achieves a reasonable trade-off between coverage and accuracy through self-consistent voting. AffectGPT-R1~\cite{AffectGPT-R1} focuses on targeted reward design during RL, promoting more effective optimization through a series of rule-based and context-driven rewards.
Along this direction, Agent-MER~\cite{Agent-MER} emulates human cognition by converting single-step emotion prediction into a structured, knowledge-driven process, achieving a balance between coverage and accuracy through self-consistent voting.
AffectGPT-R1~\cite{AffectGPT-R1} focuses on targeted reward design during RL, promoting more effective optimization.
Beyond reformulating the task itself, another line of work augments existing ones with a confidence dimension, allowing models to self-assess the acceptability of their predictions under potential subjective scenarios. EmoCaliber~\cite{EmoCaliber} progressively equips models with the ability to express and calibrate confidence, establishing a solid baseline for this direction. The methods discussed above are summarized in Tab.~\ref{tab:Understanding}.

% Subjective emotion perception arises from the fact that the same visual stimulus can elicit different emotional responses, as shown in Fig.~\ref{fig:reasoning}(b). 
% OV-MER~\cite{OVMER} elevates subjective emotion reasoning by shifting from predicting emotions within a fixed label space to an open-vocabulary, nuanced spectrum of human emotions. 
% Agent-MER~\cite{Agent-MER} presents a cognitive agent framework that converts OV-MER from a single-step process to a structured, knowledge-guided approach. 
% It employs a Knowledge-Guided Hierarchical Deliberation mechanism, based on an Emotion Tree, to systematically refine emotional understanding from coarse to fine-grained levels. 
% AffectGPT-R1~\cite{AffectGPT-R1} advances subjective emotional reasoning by employing RL to optimize emotion predictions through a set of rule-based, context-driven rewards within an OV framework. 
% EmoCaliber~\cite{EmoCaliber} verbalizes its confidence level to address subjectivity, providing users with a quantitative measure of potential interpretations while aligning with established emotion recognition frameworks.
% This approach innovatively aligns emotional reasoning with dynamic accuracy, alignment, dual, and perception rewards, enabling more nuanced and flexible emotion recognition.

\section{Model Performance Quantification} \label{sec:quantification-future}
Following a qualitative review of methods within the MER-with-LLMs paradigm, this section presents a quantitative perspective to derive further findings and insights. As shown in Tab.~\ref{tab:Results}, we observe a consistent correlation between performance and techniques across sub-tasks, revealing an evolutionary trend from freezing to fine-tuning model parameters and from SFT to RL. However, RL also introduces an observable performance fluctuation in CMER, highlighting potential room for optimization. On the other hand, methods under the Emotion Explanation and Hallucination branch tend to achieve relatively strong performance, underscoring the value of continued research into affective reasoning. 

Yet these quantitative gains may obscure several underlying concerns: many existing methods still suffer from weak zero-shot generalization, limited long-tail recognition, fragile cross-cultural transfer, sentiment hallucination, and ethical risks. Although less salient than the three major challenges discussed earlier, these issues are no less important and warrant dedicated attention in future work.

\section{Conclusion and Future Directions} \label{sec:conclusion}
%This paper presents a comprehensive survey categorizing MER with LLMs. We review recent advancements in MLLMs for MER, focusing on affective data augmentation, multimodal affective representation, and affective reasoning. For each area, we discuss representative works, their strengths, and limitations. We classify existing methods and compare key approaches both technically and experimentally. We conclude this paper with several open research directions:
% \indent These observations highlight the need for future research to move beyond performance-centric optimization and toward more principled model design and evaluation. In the following, we discuss several promising future directions for MLLMs in MER and reasoning. \\
This paper presents a comprehensive review of the emerging MER-with-LLMs paradigm. We systematically identify the key challenges and categorize existing approaches into three corresponding branches. For each branch, we discuss its motivation and representative methodologies, and compare selected works in terms of their focus, techniques, and achieved outcomes. By summarizing current progress, we aim to facilitate the future development of this paradigm. We conclude the paper by outlining the following open research directions:

\noindent
\textbf{Unified and Generalized MER.} Existing MER research is scattered across multiple sub-tasks, all of which play indispensable roles in general emotional intelligence. Therefore, research that unifies these sub-tasks is warranted, as leveraging their complementary strengths may lead to unforeseen synergies and fuel substantial vitality.

\noindent
\textbf{Mechanism-level Exploration for MER.} Understanding why often matters more than knowing what. The methodological progress of MER has left substantial gaps in mechanistic understanding, such as which model parameters critically influence emotion perception, and the underlying reasons why affective reasoning facilitates recognition. 
% and the fundamental causes behind the deficiency of general-purpose MLLMs on MER tasks.
%A deeper mechanistic understanding of how MLLMs capture emotions is essential, including the role of parameter activation, reasoning CoT, and their generalization gaps relative to standard tasks. Furthermore, developing novel training paradigms beyond SFT and RL is critical to enhance sequence modeling and supervision for nuanced emotion recognition while maintaining polarity performance.

%\textbf{Incorporating Emotional Valence in OV-MER.} OV-MER can be enhanced by integrating fine-grained emotional valence into model evaluation. Existing OV-MER largely distinguishes emotions at the synonym level, yet substantial valence differences exist between related terms, such as “content” versus “ecstatic”. Accounting for these valence variations in recognition metrics can improve sensitivity and realism, advancing nuanced understanding in MER.

\noindent
\textbf{Subjectivity-embraced Framework.} Perception subjectivity is non-negligible in the future applications of MER systems. While greatly inspiring, current frameworks still leave substantial room for refinement. For example, it remains underexplored how to establish theoretically grounded open-vocabulary emotions, as well as how to evaluate open-ended interpretations. Alternative frameworks also merit investigation, such as determining how subjectivity should be incorporated into models, and how to design subjectivity-enriched datasets and corresponding training strategies.

\noindent
\textbf{Agentic Emotion Understanding.} Direct interaction with the real world is an inevitable path for future MER development.
To bridge perception with actionable emotional intelligence, it is crucial to establish training and evaluation frameworks that support observation, planning, reasoning, tool invocation, and real-time feedback.
%The integration of agentic capabilities into MLLMs for MER remains underexplored. Unlike conventional agents that rely on externally provided tools or reasoning signals, an agentic emotion LLM would autonomously perceive emotional cues, interact with real-world multimodal environments, and dynamically retrieve relevant contextual knowledge during inference. Designing training paradigms that support planning, reasoning, and real-time tool invocation is essential for enabling robust and context-aware emotion understanding, thereby bridging perception and actionable emotional intelligence.

% \textbf{Leveraging Test-Time Scaling for Enhanced MER.} Test-time scaling (TTS) offers a promising approach to improve MLLM performance in emotion recognition. Multiple inference passes with TTS can mitigate sampling inconsistencies and yield more reliable predictions. Efficient strategies to aggregate or adapt these outputs across multimodal inputs, coupled with analysis of TTS effects on model uncertainty, are essential to further enhance robustness and accuracy in nuanced emotion recognition.

\noindent
\textbf{Safety, Bias, and Cultural Adaptation in MER.} Ensuring safety in MER requires addressing risks related to cultural variation, social bias, and individual differences. Current research largely remains at the stage of dataset construction and empirical analysis of general-purpose MLLMs, with limited exploration of concrete solutions. Advancing this area demands principled model designs and training strategies that enable culturally adaptive, bias-aware, and personalized emotion understanding for robust and responsible deployment.
% \textbf{Safety, Bias, and Cultural Adaptation in MER.} Ensuring safety in MER requires addressing risks related to cultural variation, social bias, sentiment hallucination, and personal sensitivity. Current research largely remains at the stage of dataset construction and empirical analysis of general-purpose MLLMs, with limited exploration of cross-cultural and cross-lingual settings and concrete solutions. Advancing this area demands principled model designs and training strategies that enable culturally adaptive, bias-aware, and personalized emotion understanding for robust and responsible deployment.
\section*{Acknowledgments}
% This work is supported by Beijing Natural Science Foundation (L252009), the National Natural Science Foundation of China (Nos. 62571294, 62441614), the Open Research Fund from Guangdong Laboratory of Artificial Intelligence and Digital Economy (SZ), under Grant No. GML-KF-26-09
This work is supported by Beijing Natural Science Foundation (No. L252009), the National Natural Science Foundation of China (Nos. 62571294, 62441614), the Open Research Fund from Guangdong Laboratory of Artificial Intelligence and Digital Economy (SZ) (No. GML-KF-26-09), and the Open Project of Xinjiang Key Laboratory of Multimodal Intelligent Computing and Large Models, Kashi University.
\appendix

\bibliographystyle{named}
\bibliography{ijcai26}

\end{document}